\documentclass[cmp]{svjour}
\newcommand{\R}{{\sf R\hspace*{-0.9ex}\rule{0.15ex}%
    {1.5ex}\hspace*{0.9ex}}}
\newcommand{\C}{{\sf C\hspace*{-0.9ex}\rule{0.15ex}%
    {1.5ex}\hspace*{0.9ex}}}

\begin{document}
\def \Ha #1 { {\cal H}_{ #1 } }
\def \Ho { {\Ha 1 } }
\def \Ht { {\Ha 2 } }
\def \M {M^{3n}}
\def \th {{(3)}}
\def \tw {{(2)}}
\def \a {{\alpha}}
\def \b {{\beta}}
\def \omegath {{\omega^{\th}}}
\def \omegatw {{\omega^{\tw}}}
\def \div #1 #2 {{ { #1 } \over { #2 } }}
\def \dbd #1 #2 { { \div {\partial { #1 }} {\partial { #2 }} }}
\def \pf {{\bf Proof: }}
\def \ex {{\bf Example: }}
\def \T {{\cal T}}
\def \X {{\cal X}}
\def \S {{\cal S}}
\def \Sp {{\cal SP}}
\def \Xl {{\cal X}_L}
\def \Xr {{\cal X}_R}
\def \g {{\makebox{\boldmath $g$}}}
\def \L {{\cal L }}
\def \RR {{\cal R}}
\def \s {{\makebox{\boldmath $\sigma$}}}

\spnewtheorem{defn}{Definition}{\bf}{\it}
\spnewtheorem{thm}{Theorem}{\bf}{\it}
\spnewtheorem{prop}{Proposition}{\bf}{\it}
\spnewtheorem{cor}{Corollary}{\bf}{\it}

\title{ Generalized Nambu system on $S^3$ and spinors} 
\author{ Sagar A. Pandit\inst{1}\and Anil D. Gangal\inst{2}}
\institute{Physical Research Laboratory, Navrangpura, Ahmedabad 380
  009, India\\ \email{sagar@prl.ernet.in} \and Department of
  Physics, University of Pune, Pune 411 007, India.\\ \email{adg@physics.unipune.ernet.in}}
\communicated{Sagar A. Pandit}
\maketitle
\begin{abstract}
It is shown that the manifold $S^3$ can be equipped with a natural
Nambu structure arising out of a cross product on the tangent
space. Further, the group action of $SU(2)$ on $S^3$ is shown to be Nambu
action. Moreover, we compare the action of $SU(2)$ on spinors with
that of a Nambu system.
\end{abstract}

\section{ Introduction }
There exist certain dynamical systems which resist the usual canonical
description~\cite{AMR,PJM}. Nambu dynamics and its various
generalizations~\cite{Nambu,Mukunda,BAndF,KAndT,Takht,Esta,Fecko}
offer an appropriate possibility for the description of a certain class
of non-canonical dynamical systems.

 In~\cite{Sag} we proposed a geometric formulation of generalized Nambu
systems. The motivation behind that work was to provide a framework
suitable from the point of view of dynamical systems. The framework proposed
in~\cite{Sag} involves a $3n$ dimensional manifold $\M$, together with
a closed, strictly  non-degenerate  3-form $\omegath$. The time
evolution is governed by a pair of Nambu functions $\Ho, \Ht$.  It was
seen that the formulation involving 2-forms provides a natural
approach to a Nambu system. The connection between
symmetries of Nambu systems and conserved 2-forms is established
in~\cite{Sag1} by introducing the  notion of  a ``Nambu momentum map''. A 
generalization of the symplectic Noether theorem for Nambu systems is
carried out in~\cite{Sag1}.  

In case of Hamiltonian systems the manifold $\R \times T^1$ plays a
special role. We expect that the manifold $S^3$ should play similar
role for Nambu systems. The purpose of the present paper is two
fold. Firstly, we explicitly construct a Nambu system on a
non-trivial (non vector space) manifold $S^3$ using the cross
product. Secondly, we study the Nambu action of $SU(2)$ on $S^3$ and 
identify  that the action of $SU(2)$ on the normalized spinor is
equivalent to the action of natural Nambu vector field on $S^3$. The
relevance of this observation is obvious.

In section~\ref{NambuSystems} we list the main features of Nambu
systems developed in~\cite{Sag,Sag1}. In section~\ref{NambuS3}
we give natural Nambu structure on $S^3$ and show the equivalence
between the action of Nambu vector field on $S^3$ and action of
$SU(2)$ on normalized spinors.. 

\section{ Nambu Systems }
\label{NambuSystems}
In this section we briefly review the relevant definitions and
results regarding Nambu systems~\cite{Sag,Sag1}. 

\begin{defn} (Nambu Manifold) : Let $\M$ be a $3n$-dimensional $C^\infty$
  manifold and let $\omegath$ be a 3-form field on $\M$ such that
  $\omegath$ is completely anti-symmetric, closed and strictly
non-degenerate (See~\cite{Sag}) at every
  point of $\M$ then the pair $(\M,\omegath)$ is called a
  {\em Nambu manifold}. 
\end{defn}

 The Nambu structure preserving transformation (Nambu canonical
 transformation) is defined in~\cite{Sag,Sag1}. Using this, a
 generalization of Darboux theorem is proved, which gives a
 Nambu-Darboux coordinates on Nambu manifold~\cite{Sag,Sag1}.
 We define the analogs of raising and lowering operations.
The map $\flat : \X(\M) \rightarrow \Omega^0_2(\M)$ is defined by $ X
\mapsto X^\flat = i_X \omegath$, where $\X(\M)$ is space of vector
fields on $\M$ and $\Omega^0_2(\M)$ is space of 2-forms on $\M$. The
map $\sharp : \Omega^0_2(\M) \rightarrow \X(\M)$, is defined by the
following prescription.  Let $\alpha$ be a 2-form and $\alpha_{ij}$ be its
components in Nambu-Darboux coordinates, then the components of
$\alpha^\sharp$ are given by~\cite{Sag,Sag1}
\begin{eqnarray} 
{\alpha^\sharp}^{3i+p} = {1 \over 2} \sum_{l,m=1}^3 \varepsilon_{plm}
\alpha_{3i+l \;\;\; 3i+m}  \nonumber
\end{eqnarray}
These maps provide a bracket structure of 2-forms on Nambu
manifold~\cite{Sag,Sag1} as, let $\alpha, \beta \in \Omega^0_2(\M)$,
then the bracket is $\{ \alpha, \beta\} = [ \alpha^\sharp,
\beta^\sharp]^\flat$, where $[\;\;,\;\;]$ is Lie bracket of vector
fields~\cite{Sag,Sag1}. 

The Nambu action and the Nambu momentum maps are defined
in~\cite{Sag1} as
\begin{defn} (Nambu action) : Let $G$ be a Lie group and let
  $(\M,\omegath)$ be a Nambu manifold. Let $\Phi_G$ be the action of $G$ on
  $\M$. $\Phi_G$ is called {\em Nambu action} if 
  \begin{eqnarray}
    \Phi_G^* \omegath = \omegath \nonumber
  \end{eqnarray}
  i.e., $\Phi_G$ induces a Nambu canonical transformation.
\end{defn}

A quantity similar to symplectic momentum map is defined in~\cite{Sag1}.
\begin{defn} (Nambu momentum maps) : Let $G$ be a Lie group and let
  $(\M, \omegath)$ be a Nambu manifold, let $\Phi_G$ be a Nambu action on
  $\M$. Then the mapping ${\bf J} \equiv (J_1, J_2) : \M \rightarrow
  \g^* \times \g^*$ is called {\em Nambu momentum maps} provided,
  for every $\xi \in \g$ 
  \begin{eqnarray}
    (d\hat{J}_1 (\xi) \wedge d\hat{J}_2 (\xi))^\sharp = \xi_{\M} \nonumber
  \end{eqnarray}
  where $\hat{J}_1(\xi) : \M \rightarrow \R$, $\hat{J}_2 (\xi) :
  \M\rightarrow\R$, defined by $\hat{J}_1 (\xi)(x) = J_1(x) \xi$ and
  $\hat{J}_2(\xi)(x)= J_2(x)\xi$ $\forall x \in\M$ and $\xi_{\M}$ is
  infinitesimal generator of the action $\Phi_G$.
\end{defn} 

\begin{defn} (Nambu G-space) : The five tuple
  $(\M,\omegath,\Phi_G,J_1,J_2)$ is called {\em Nambu G-space}. 
\end{defn}

The consistency between the bracket of 2-forms and the Nambu momentum 
maps follows from the proposition established in~\cite{Sag1} viz
\begin{prop}\label{Consistency} Let $(\M,\omegath,\Phi_G, J_1, J_2)$ be
  a Nambu G-space and $\xi,\eta \in \g$ then 
  \begin{eqnarray}
    (d\hat{J}_1([\xi,\eta]) \wedge d\hat{J}_2([\xi,\eta]))^\sharp = \{
    d\hat{J}_1(\eta) \wedge d\hat{J}_2(\eta), d\hat{J}_1(\xi) \wedge
    d\hat{J}_2(\xi) \}^\sharp 
    \nonumber
  \end{eqnarray}
i.e., The following diagram commutes \\
\begin{center}
\begin{picture}(100,100)
\put(5,80){$\Omega^0_2(\M)$}
\put(145,80){$\X(\M)$}
\put(5,10){$\g$}
\put(70,87){$\sharp$}
\put(-17,40){$\hat{J_1}, \hat{J_2}$}
\put(72,40){$\xi \mapsto \xi_{\M}$}
\thicklines
\put(50,82){\vector(1,0){90}}
\put(9,17){\vector(0,1){58}}
\put(11,17){\vector(2,1){129.4}}
\end{picture}
\end{center}
\end{prop}

In analogy with the notion of symplectic symmetry for the Hamiltonian
system, the idea of Nambu Lie symmetry is introduced in~\cite{Sag1}. 
\begin{defn} (Nambu Lie symmetry) : Consider a Nambu system
  $(\M,\omegath,\Ho,\Ht)$. Let 
  $(\M,\omegath,\Phi_G,J_1,J_2)$ be a Nambu G-space. We call $\Phi_G$ a
  {\em Nambu Lie symmetry} of the Nambu system if
  \begin{eqnarray}
    \Phi_G^* (d\Ho \wedge d\Ht) = (d\Ho \wedge d\Ht) \nonumber
  \end{eqnarray}
\end{defn}

The Noether theorem in this framework now reads as~\cite{Sag1}
\begin{thm} \label{Noether}Consider a Nambu system $($ $\M, \omegath,
\Ho, \Ht$ $)$ where 
  $\Ho, \Ht$ are so chosen that $d\Ho \wedge d\Ht = {(d\Ho \wedge
    d\Ht)^\sharp}^\flat$. Let this system be a Nambu G-space
  $(\M,\omegath,\Phi_G,J_1,J_2)$  where $J_1, J_2$ are so chosen that
  $d\hat{J}_1(\xi) \wedge d\hat{J}_2(\xi) = {(d\hat{J}_1(\xi) \wedge
    d\hat{J}_2(\xi))^\sharp}^\flat$ $\forall \xi \in \g$. 
  If $\Phi_G$ is Nambu Lie symmetry of this system
  then
  \begin{eqnarray}
    L_{(d\Ho \wedge d\Ht)^\sharp} \Big(d\hat{J}_1(\xi) \wedge d\hat{J}_2(\xi)\Big)
    = 0 \nonumber 
  \end{eqnarray}
  i.e., $d\hat{J}_1(\xi) \wedge d\hat{J}_2(\xi)$ is conserved by  the
  Nambu flow.
\end{thm}

\section{ Spinors and the Nambu system on $S^3$ }
\label{NambuS3}
The manifold $\R\times T^1$ plays a special role in Hamiltonian
systems. We expect that $S^3$ should play a similar role in
case of Nambu systems.  Here, we show that $S^3$ can be equipped with
a Nambu structure, thus provide a non-trivial (other than $\R^{3n}$)
example of Nambu manifold. We also demonstrate that the action of
$SU(2)$ on $S^3$ is a Nambu action and notice the similarity with 
evolution of Pauli spinors.

\subsection{ $SU(2)$ Spinors}

The group $SU(2)$ is a group of complex unimodular unitary $2\times2$
matrices. A standard parameterization of an element of $SU(2)$ is
\begin{eqnarray}
U_{\makebox{ {\boldmath $n$}}}(\theta) = \exp({ {-i\theta} \over
2} {\makebox{\boldmath $n$}} \cdot \s)  = \makebox{\boldmath $1$}
\cos({\theta \over 2}) + i\;\;\;  {\makebox{\boldmath $n$}} \cdot \s
\sin({\theta \over 2}) \nonumber 
\end{eqnarray}
where ${\makebox{\boldmath $n$}}$ is a unit vector and
\begin{eqnarray}
\sigma_1 = \left(\begin{array}{cc} 0 & 1 \\ 1 & 0\end{array}\right),
\sigma_2 = \left(\begin{array}{cc} 0 & -i \\ i & 0\end{array}\right),
\sigma_3 = \left(\begin{array}{cc} 1 & 0 \\ 0 & -1\end{array}\right)
\nonumber
\end{eqnarray}
are the {\em Pauli matrices}. The elements of $\C^2$ are the $SU(2)$
spinors. 

For various reasons and especially in view of its relevance in quantum
mechanics the action of $SU(2)$ on the normalized spinors is of great
interest (Here the normalization is carried out with the hermitian
scalar product). We denote the space of normalized spinors by $\Sp$. \\

\noindent
{\bf Remarks:} \\
The space of normalized spinors is $S^3$. \\
Let $\xi \in \Sp$. So without loss of generality we write 
\begin{eqnarray}
\xi = \left( \begin{array}{c} x + i y \\ z + i w \end{array} \right)
\nonumber
\end{eqnarray}
where $x,y,z,w \in \R$ and the condition $\xi^\dagger\xi = 1$ implies 
that $x^2 + y^2 + z^2 + w^2 = 1$, which is $S^3$

\subsection{Nambu system on $S^3$}
\label{NambuSystemOnS3}
\begin{prop} \label{Nambu-equipped} The manifold $S^3$ is equipped
with a natural Nambu structure. 
\end{prop}
\pf Since  $S^3$ is the group manifold of the Lie group $SU(2)$ we
define the 3-form at the identity and the by left action on can define
it everywhere.

The Lie algebra of $SU(2)$ is $su(2)=\R^3$. Every three dimensional
space has vector cross product~\cite{Marathe}. So we define 3-form in $T_e S^3$ as 
\begin{eqnarray}
  \epsilon(a,b,c) = \omegath(e)(a,b,c) = (a, b \times c) \;\;\; where
  a,b,c \in T_eS^3 
  \nonumber
\end{eqnarray}
The 3-form $\epsilon$ is non-degenerate and hence
strictly non-degenerate since the space is three dimensional~\cite{Sag}.

Now we define $\omegath(g) = {\cal L}^*_{g^{-1}} \epsilon$, where
${\cal L}_g$ is the left action corresponding to $g\in SU(2)$, as the
3-form every where. Since $d\omega = 0$, the form $\omegath$ is also
closed. So we have the Nambu structure $\omegath$ on $S^3$.
\newline\qed

Since the 3-form $\omegath$ is defined by the left action it is
natural to expect that the action is a Nambu action.
\begin{prop} The left action of $SU(2)$ on $S^3$ is Nambu action.
\end{prop}
\begin{eqnarray}
  {\cal L}_h^* \omegath(g) &=& {\cal L}^*_h {\cal L}^*_{g^{-1}} \epsilon
  \nonumber \\
  &=& {\cal L}^*_{hg^{-1}}\epsilon \nonumber \\
  &=& \omegath(h^{-1}g) \nonumber
\end{eqnarray}
By proposition~\ref{Nambu-equipped}, the left action is a Nambu action.
\newline\qed

We construct Momentum maps corresponding to this action. Lets
consider the Lie algebra of $SU(2)$ which is $su(2) = \g =
\R^3$. Let $e_1, e_2, e_3$ be a basis of $\g$ satisfying the following 
bracket conditions. 
\begin{eqnarray} 
  [ e_1, e_2 ] = 2e_3,\;\;
  [ e_2, e_3 ] = 2e_1,\;\;
  [ e_3, e_1 ] = 2e_2 \nonumber
\end{eqnarray}
Let $f_1, f_2, f_3$ be a dual basis of $\g^*$ corresponding to $e_1,
e_2, e_3$. We choose a {\em Nambu-Darboux} coordinate system on $S^3$. Let
$p\in S^3$ be represented by $(x,y,z)$ in this coordinate system. 
Let $\vec{r} \equiv x f_1 + y f_2 + z f_3, \vec{\rho} \equiv (y^2 +
z^2) f_1 +  (x^2 + z^2) f_2 +  (y^2 + x^2) f_3 \in \g^*$. Then
the momentum maps are $J_1(x,y,z) = \vec{r}$ and $J_2(x,y,z) =
\vec{\rho}$. The 2-forms obtained from these are 
\begin{eqnarray}
  d\hat{J}_1(e_1) \wedge d\hat{J}_2(e_1) &\equiv& S_1 = 2y d x \wedge d y + 2z d x
  \wedge d z \nonumber \\ 
  d\hat{J}_1(e_2) \wedge d\hat{J}_2(e_2) &\equiv& S_2 = - 2x d x \wedge d y + 2z
  d y \wedge d z \nonumber \\ 
  d\hat{J}_1(e_3) \wedge d\hat{J}_2(e_3) &\equiv& S_3 = - 2x d x \wedge d z - 2y
  d y \wedge d z \nonumber 
\end{eqnarray}
Corresponding to these 2-forms we write vector fields. By the definition
of momentum  maps we write $S_1 = i_{S_1^\sharp} \omegath =
i_{S_1^\sharp} (dx \wedge dy \wedge dz) \Rightarrow S_1^\sharp \equiv
(0, -2z, 2y)$
Similarly
\begin{eqnarray}
  S_2^\sharp &=& (2z,0,-2x) \nonumber \\
  S_3^\sharp &=& (-2y,2x,0) \nonumber
\end{eqnarray}
Using the definition of the Nambu bracket, clearly, $\{ S_1, S_2 \} =
i_{[S_1^\sharp, S_2^\sharp]} \omegath = L_{S_1^\sharp} i_{S_2^\sharp}
\omegath - i_{S_2^\sharp} L_{S_1^\sharp} \omegath$. Since $\omegath,
S_1, S_2, S_3$ are closed forms, $\{ S_1, S_2 \} = d
(i_{S_1^\sharp} S_2) = -2S_3$. Similarly
\begin{eqnarray} 
\{ S_2, S_3\} = - 2S_1, \{S_3, S_1 \} = - 2S_2. \label{SAlgebra}
\end{eqnarray}
This establishes the consistency condition stated in
Proposition~\ref{Consistency}.

As shown above the left action of $SU(2)$ is Nambu action. Which
implies that there are momentum maps which are generators of such
an action. From equation(~\ref{SAlgebra}) it is clear that the algebra of
the generators of Nambu action is isomorphic to the algebra of Pauli
spin matrices. 

\subsection{ Spin systems and Nambu systems}

In section~\ref{NambuSystemOnS3}, we have shown that the action of
$SU(2)$ on $S^3$ is Nambu action. In this section, through
Proposition~\ref{equivalence}, we show that such a Nambu action is
equivalent to the action of $SU(2)$ on normalized spinors.

\begin{prop} \label{equivalence} Let $h : S^3 \rightarrow \Sp$
defined by\footnote{ Since $S^3$ is group manifold of group $SU(2)$we
are not distinguishing between a point in $SU(2)$ and a point of $S^3$}
$\left(\begin{array}{cc} x + i y & -z + iw \\ z + i w & x - iy
\end{array} \right) \mapsto \left( \begin{array}{c} x + iy \\ z + iw
\end{array} \right)$ where $x,y,z,w \in \R^4$ satisfying condition
$x^2 + y^2 + z^2 + w^2 = 1$. Let $\Phi_U$ be the induced action of
$SU(2)$ on $\Sp$. Then the Nambu action of $SU(2)$ on $S^3$ is
equivalent to the action $\Phi_U$ of $SU(2)$ on $\Sp$. i.e. The
following diagram commutes.\\
\begin{center}
\begin{picture}(100,100)
\put(5,80){$S^3$}
\put(95,80){$\Sp$}
\put(5,10){$S^3$}
\put(95,10){$\Sp$}
\put(50,85){$h$}
\put(50,15){$h$}
\put(-7,50){$\L_g$}
\put(105,50){$\Phi_U$}
\thicklines
\put(9,77){\vector(0,-1){57}}
\put(17,12){\vector(1,0){75}}
\put(17,82){\vector(1,0){75}}
\put(100,77){\vector(0,-1){57}}

\end{picture}
\end{center}
\end{prop}
\pf Let $\xi_M = \left(\begin{array}{cc} x + i y & -z + iw \\ z + i w & x - iy
\end{array} \right) \in S^3$. The left action of $SU(2)$ with the
above parameterization takes this point to 
$\xi^\prime_M = \left(\begin{array}{cc} x^\prime + i y^\prime & -z^\prime + iw^\prime
\\ z^\prime + i w^\prime & x^\prime - iy^\prime \end{array} \right)$
where
\begin{eqnarray}
x^\prime &=& x \cos({\theta \over 2}) - y\; n_3 \sin({\theta \over 2}) + z\;
n_2 \sin({\theta \over 2}) - w\; n_1 \sin({\theta \over 2}) \nonumber \\ 
y^\prime &=& x\; n_3 \sin({\theta \over 2}) + y \cos({\theta \over 2}) + z\;
n_1 \sin({\theta \over 2}) + w\; n_2 \sin({\theta \over 2}) \nonumber \\ 
z^\prime &=& - x\; n_1 \sin({\theta \over 2})  - y\; n_2 \sin({\theta \over
2}) + z \cos({\theta \over 2}) - w\; n_3 \sin({\theta \over 2})
\nonumber \\
w^\prime &=& x\; n_2 \sin({\theta \over 2}) - y\; n_1 \sin({\theta \over 2})
- z\; n_3 \sin({\theta \over 2}) + w \cos({\theta \over 2}) \nonumber 
\end{eqnarray}
Now
\begin{eqnarray}
\Phi_U h (\xi_M) = U_{\makebox{ {\boldmath $n$}}}(\theta) \left(
\begin{array}{c} x + iy \\ z + iw \end{array} \right) = \left(
\begin{array}{c} x^\prime + iy^\prime \\ z^\prime + iw^\prime
\end{array} \right) \equiv \xi^\prime \nonumber
\end{eqnarray}
which is
\begin{eqnarray}
\xi^\prime = h (\xi^\prime_M) \nonumber
\end{eqnarray}
\newline\qed\\

\noindent
{\bf Remarks:}
\begin{enumerate}
\item Consider quantum mechanical evolution of a charged spin ${1 \over 2}$
particle in the presence of a constant uniform external magnetic
field. The time evolution of normalized Pauli spinor can be considered
as action of $SU(2)$ on the spinor. Hence, in view of
Proposition~\ref{equivalence} the time evolution can be considered as
Nambu evolution. 
\item Classical pure spin systems are considered in Hamiltonian
  framework~\cite{Sudarshan}. Proposition~\ref{equivalence} suggests
  that Nambu framework with generalized Nambu-Poisson bracket is
  a suitable framework for classical pure spin systems.
\end{enumerate}

\section{ Conclusions }
In this paper we have constructed a Nambu system on $S^3$. Further the 
action of $SU(2)$ on such a system is identified as Nambu
action. In fact, this observation has an obvious relevance for the time 
evolution of Pauli spinor in quantum mechanics. We hope that eventually 
such construction will find their use in construction of spin
manifold~\cite{Milnor,Chichilnsky}.

\begin{acknowledgement}
We thank Dr. H. Bhate for discussion.
\end{acknowledgement}


\end{document}